# NDUI+: A fused DMSP-VIIRS based global normalized difference urban index dataset

Manmeet Singh[1,2]**, Subhasis Ghosh[3]**, Harsh Kamath[1], Shivam Saxena[4], Vaisakh SB[2], Chandana Mitra[3], Naveen Sudharsan[1], Suryachandra Rao[2], Hassan Dashtian[5], Lori Magruder[6], Marshall Shepherd[7], Dev Niyogi[1,6,8]*

[1] Jackson School of Geosciences, The University of Texas at Austin, Austin, Texas, United States
[2] Indian Institute of Tropical Meteorology, Ministry of Earth Sciences, Pune, India
[3] Department of Geosciences, Auburn University, Auburn, Alabama, United States
[4] National Institute of Technology, Rourkela, India
[5] Bureau of Economic Geology, The University of Texas at Austin, Austin, TX, United States
[6] Centre for Space Research, The University of Texas at Austin, Austin, TX, United States
[7] Department of Geography, University of Georgia, Athens, Georgia, United States
[8] Maseeh Department of Civil, Architectural, and Environmental Engineering, The University of Texas at Austin, Austin, TX, United States

* Correspondence to dev.niyogi@jsg.utexas.edu (Dev Niyogi)
**M. Singh and S. Ghosh are equal contributors to this work and are designated as co-first authors

**Abstract**

Urbanization is advancing rapidly, covering less than 2% of Earth's surface yet profoundly influencing global environments and experiencing disproportionate impacts from extreme weather events. Effective urban management and planning require high-resolution, temporally consistent datasets that capture the complexity of urban growth and dynamics. This study presents NDUI+, a novel global urban dataset addressing critical gaps in urban data continuity and quality. NDUI+ integrates data from the Defense Meteorological Satellite Program's Operational Linescan System (DMSP-OLS), VIIRS Nighttime Light, and Landsat 7 NDVI using advanced remote sensing and deep learning techniques. The dataset resolves sensor discontinuity challenges, offering a seamless 30-meter spatial and annual temporal resolution time series from 1999 to the present. NDUI+ demonstrates high precision and granularity, aligning closely with high-resolution satellite data and capturing urban dynamics effectively. The dataset provides valuable insights for urban climate studies, IPCC assessments, and urbanization research, complementing resources like UT-GLOBUS for urban modeling.

## 1. Introduction

This study introduces an innovative methodology to develop a fused multidecadal, high-resolution global normalized difference urban index (NDUI+) dataset by integrating DMSP-OLS and VIIRS nighttime light data using advanced deep learning techniques. The NDUI+ dataset, created with the Swin Transformer model, provides a robust framework for understanding urbanization processes at both global and local scales, addressing a critical gap in urban monitoring caused by the discontinuation of DMSP-OLS data.

Urbanization continues to rise rapidly, with cities now hosting more than half of the global population (World Bank, 2023). Despite occupying less than 2% of the Earth's surface, urban areas account for over 70% of carbon dioxide emissions from energy use (Seto et al., 2012) and play a significant role in shaping global climate and environmental feedbacks (Okie, 1986; Liu & Niyogi, 2017; Sui et al., 2024). Urbanization drives profound changes in local climates, influencing temperature, precipitation, and the broader socio-environmental landscape. These impacts are accompanied by rapid population growth, economic development, and increasing spatial expansion, underscoring the need for precise, high-resolution data to assess urbanization dynamics and develop sustainable management strategies (Dodman et al., 2022).

Understanding the multifaceted impacts of urbanization requires long-term, high-resolution datasets that capture urban patterns across scales. Remote sensing, a critical tool for urban analysis, provides a wealth of Earth observation data at varying spatiotemporal resolutions, enabling the monitoring of urban growth, land use, and environmental change (Netzband et al., 2007). Despite the availability of various datasets such as the World Settlement Footprint (Marconcini et al., 2020), GHSL Settlement Grid (Pesaresi et al., 2016), and Global Urban Footprint (Santangelo et al., 2022), existing indices often lack the temporal continuity and spatial granularity necessary for capturing dynamic urban changes over decades. The NDUI index, developed to assess impervious surfaces and human settlement patterns, previously filled this gap but became obsolete after 2014 due to the decommissioning of DMSP satellites (Guo et al., 2017; Ma et al., 2018; Wang et al., 2022). This study addresses this limitation by developing a new, AI-driven methodology to ensure data continuity and improve urban monitoring capabilities.

The Swin Transformer, a novel vision transformer model, forms the foundation of this methodological advancement. By employing non-overlapping, displaced local windows, the Swin Transformer efficiently captures both local and global image features, enabling the seamless fusion of DMSP-OLS and VIIRS datasets (Liu et al., 2021). This deep learning approach allows the reconstruction of DMSP-OLS-like data from VIIRS inputs, eliminating the dependency on outdated datasets while maintaining compatibility with next-generation satellite technologies. The resulting NDUI+ dataset spans 1999 to the present, providing 30-meter resolution data with annual updates. This temporal and spatial consistency makes NDUI+ particularly suitable for analyzing urban-climate interactions, urban resilience, and environmental change.

Urban datasets serve a wide array of applications, including socioeconomic assessments (Jiang et al., 2017), city planning (Madlener & Sunak, 2011), environmental studies (Xu, 2010), mobility research (Smith, 2020), and climate modeling (Chapman et al., 2017). NDUI+ enhances these capabilities by offering fine-scale, multidecadal data for detailed urban analysis. Its applications include mapping impervious surfaces (Guo et al., 2017), estimating human settlements (Ma et al., 2018), and identifying urban infrastructure changes (Xu et al., 2023). The dataset's ability to delineate urban boundaries while capturing intra-urban details makes it invaluable for studies on urban morphology, climate feedbacks, and sustainability.

This study directly advances the understanding of urban dynamics through the development of a high-resolution, multidecadal annual urban data capable of identifying fine features in urban areas. By bridging generational gaps in satellite data, NDUI+ enables researchers to analyze urbanization patterns and their socio-environmental impacts over time, providing actionable insights for cities in diverse geographic and socio-economic contexts. This novel approach facilitates interdisciplinary collaboration, empowering urban planners, policymakers, and scientists to address critical challenges in urban development, resilience, and climate adaptation. NDUI+ stands as a transformative tool for global urban studies, fostering a deeper understanding of cities as complex, evolving systems.

| Dataset | Spatial resolution / grid size | Period | Temporal resolution | Global |
|---|---|---|---|---|
| Global Impervious Surface Area (Huang et al., 2021) | 10 m | 2015 and 2018 | Two Static maps | ✓ |
| National Land Cover Database[a] urban imperviousness (NLCD, Dewitz et al. 2021) | 30 m | 2001 to 2021 | Every 3 years | x |
| Global Urban Footprint (GLUE, Santangelo et al., 2022) | 30 m | 2020 | One time (Static) | ✓ |
| Global urban growth (Li et al., 2021) | 1000 m | 1870 to 2100 | 10 years | ✓ |
| Global Urban land expansion (Chen et al., 2020) | 1000 m | 2015 to 2100 | 10 years | ✓ |
| Global Impervious Surface Dynamic Dataset (Zhang et al., 2022) | 30 m | 1985 to 2020 | 5 years | ✓ |
| Global Intra-Urban Land Use[b] (Guzder-Williams et al., 2023) | 5 m | 2020 | One time (Static) | ✓ |
| DynamicWorld (Brown et al. 2022) | 10 m | 2015-present | ~6-15 days | ✓ |
| NDUI+ data (Proposed by this study) | 30 m | 1999-Present | 1 year | ✓ |

Table 1: Summary of the different datasets available for characterizing and studying urban areas. [a]NLCD is not global and is only for US. [b]This data has a 'binary' urban/no urban index. All other datasets have urban fraction in truncation. NDUI+ is additionally able to granualarize within urban areas.

## 2. Results

### 2.1 Urban Land Use Detected by NDUI+ Data

This study successfully developed a multidecadal (1999–present), spatiotemporally continuous (yearly), high-resolution (30 m) global urban dataset, NDUI+, offering unprecedented potential for monitoring urban patterns at scales ranging from local to global. The dataset enables detailed observation of urban expansion, land use changes, and built environment dynamics over time. For example, Figure 1 illustrates urban expansion in Austin, Texas, USA, as detected by NDUI+. The high spatial and temporal resolution of NDUI+ makes it particularly effective in capturing the evolution of urban land use patterns in fast-growing cities, emphasizing its value for studies focusing on urbanization processes and their socio-environmental implications.

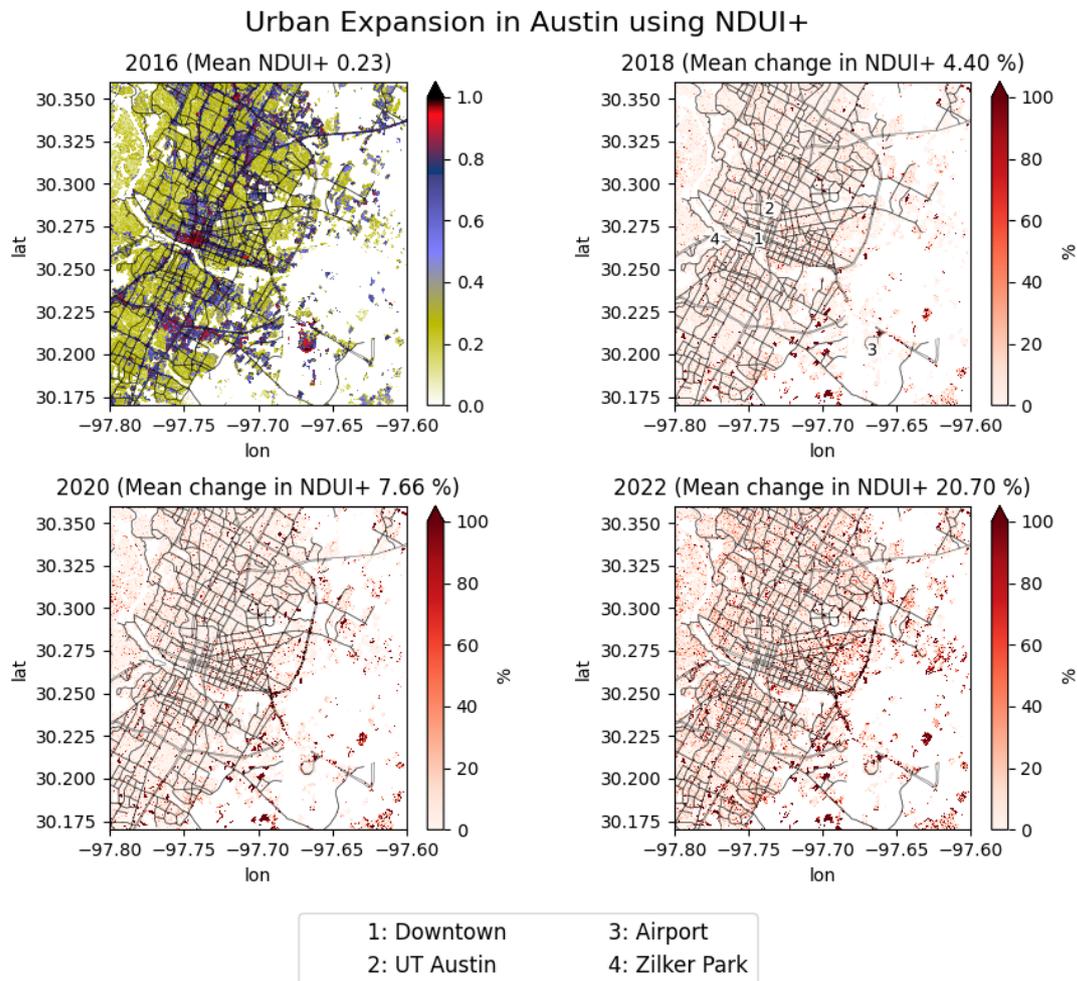

*Figure 1: Austin, Texas, USA over time as detected by NDUI+. (a) shows NDUI+ values for 2016 whereas (b-d) subplots represent percentage change in NDUI+ relative to 2016 for the years 2018, 2020 and 2022, respectively. In this figure, NDUI+ is shown as filtered by Dynamic World built up averaged for that particular year. The pixels with Dynamic World built up class with values less than 0.2 for the year 2023 are masked out.*

### *2.2 Enhancing DMSP-OLS Reconstruction Using Spatial Fine-Tuning*

To address the limitations of earlier DMSP-OLS datasets, a pixel-by-pixel comparison was performed between the extended DMSP-OLS data generated in this study (Dataset A) and the version developed by Ghosh et al. (2021) (Dataset B), using a time series for Las Vegas, USA. Both datasets indicated a decline in average radiance post-2012, aligning with the Las Vegas Home Price Index (https://fred.stlouisfed.org/series/LVXRNSA), reflecting reduced urban activity. However, Dataset A showed recovery starting in 2016, consistent with local economic recovery (Figure S1). In contrast, Dataset B lacked this recovery, likely due to its reliance on generalized model training without regional specificity.

Further validation was performed by training the model with data from Austin, Texas, and then applying it to simulate DMSP-OLS data for other cities. Both datasets produced similar outcomes, confirming model consistency. However, when trained specifically on Las Vegas data, Dataset A displayed enhanced accuracy, particularly in depicting post-2016 trends. These findings highlight the heterogeneity of urban systems and underscore the importance of region-specific model calibration to accurately simulate urban dynamics. This approach emphasizes the need for localized methodologies when interpreting urban data across diverse geographies and socio-economic contexts.

*2.3 NDUI+ Data for Diverse Urban Environments Across the United States*

The NDUI+ dataset provides yearly updates, enabling detailed temporal analysis of urban development. A comparison of urban growth trends in nine U.S. cities—Albuquerque (New Mexico), Austin (Texas), Columbus (Ohio), Kansas City (Kansas), Las Vegas (Nevada), Minneapolis (Minnesota), Portland (Oregon), Seattle (Washington), and Washington D.C.—is presented in Figure 2. These cities were selected to represent a wide spectrum of urban environments, from arid climates like Albuquerque and Las Vegas to wet, temperate regions like Portland and Seattle. NDUI+ effectively captures the nuances of urban growth and provides a granular understanding of yearly urban changes. Its ability to track development across diverse climatic and geographic conditions demonstrates its applicability for nationwide urban studies and cross-comparative analyses.

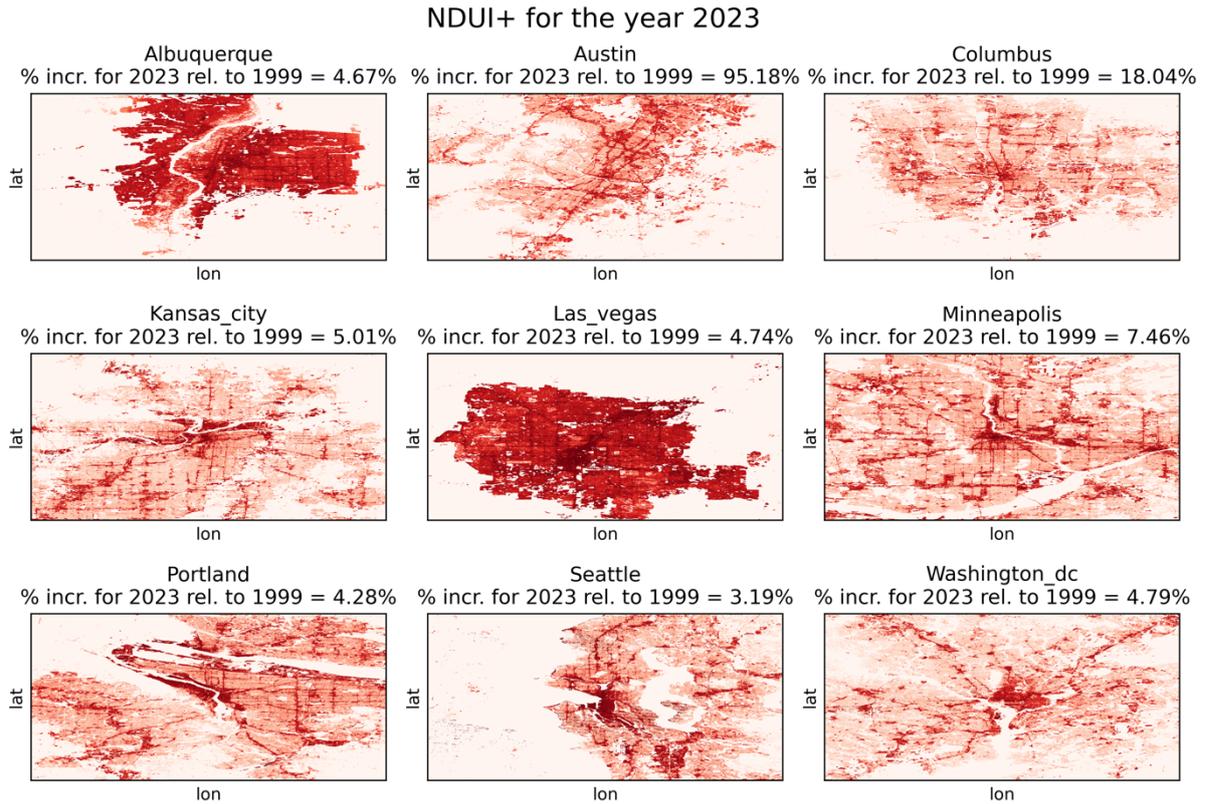

*Figure 2: NDUI+ for different US cities for the year 2023. Subtitles show the percentage increase in NDUI+ for the year 2023 relative to 1999.*

## 2.4 Comparison of NDUI+ with Sentinel-2-Based Dynamic World Built-Up Dataset

To evaluate the accuracy and granularity of NDUI+, it was compared with the Sentinel-2-based Dynamic World built-up dataset for a neighborhood in Austin, Texas (Figure S2). This location was chosen for its accessibility and suitability for ground-truth validation. Due to differences in dataset characteristics, relative rather than absolute values were used for the analysis. The results show that NDUI+ effectively distinguishes urban features, such as roads in Oakwood Cemetery and low-built-up areas in Hempfield Park, demonstrating high granularity in detecting urbanization levels.

A detailed analysis of a stadium revealed that Dynamic World classified the entire stadium as a high built-up area, while NDUI+ differentiated the grass-covered central field, assigning it a lower urbanization value. Both datasets successfully identified smaller features, such as the UT Tower, but NDUI+ outperformed in identifying low-valued areas surrounding the tower (Figure 3). Additionally, NDUI+ displayed high fidelity in detecting fine-scale urban features such as green roofs within dense urban neighborhoods (Figure 4). These results underscore NDUI+'s suitability for local climate zone analysis, urban climate modeling, and infrastructure assessments at both city and street scales.

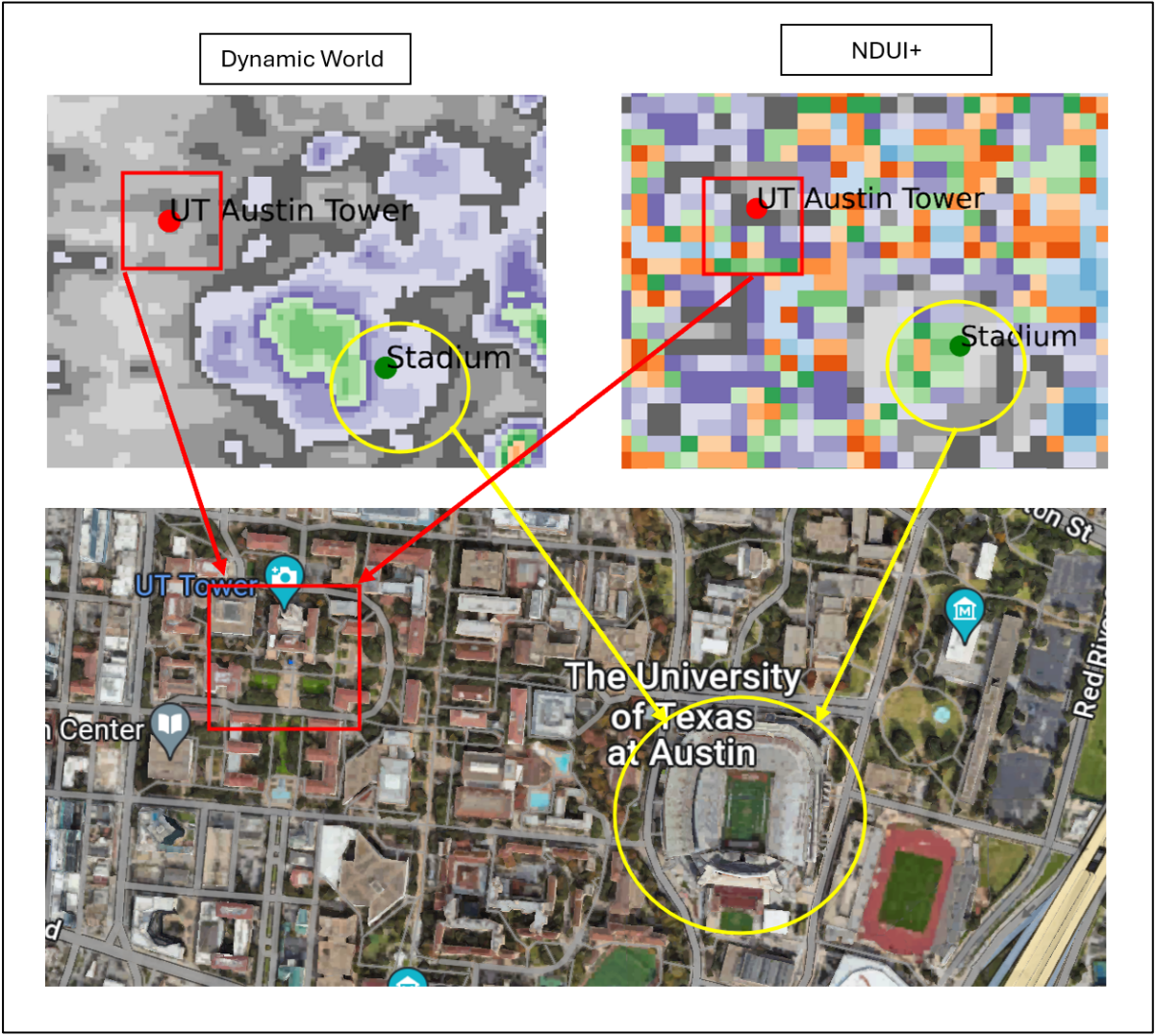

*Figure 3: The left plot in the top panel shows Dynamic world built up dataset, The top right shows NDUI+ data from this study and the bottom panel shows the corresponding 'ground truthing' using Google Earth imagery. Example of NDUI+ ability to differentiate macro urban landscapes such as built up area (buildings), greenspaces (stadiums).*

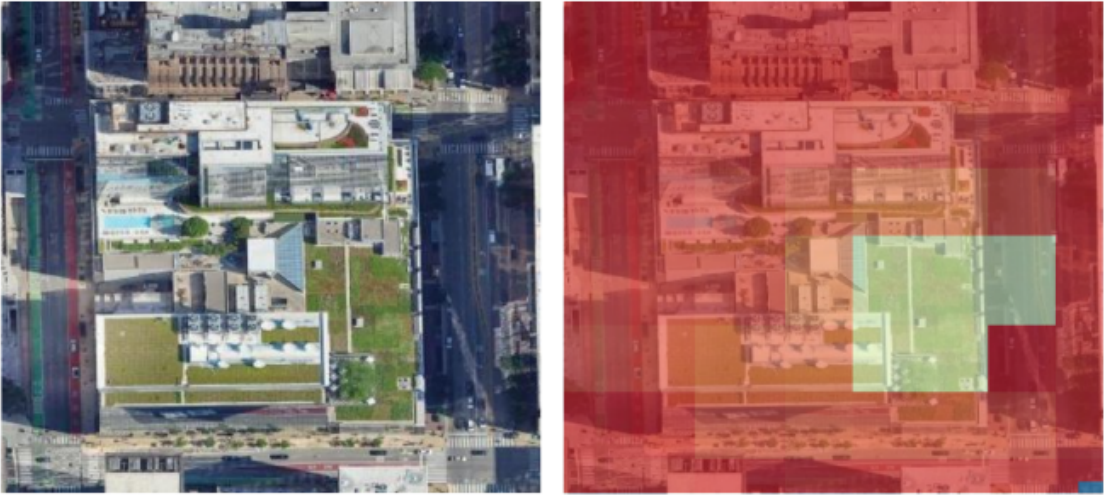

*Figure 4: Example showing NDUI+ identifying microscale granular features such (1) Roads, (2) green roof in a (3) dense concrete urban area. The example shown is of Chicago downtown.*

## 2.5 Urban Land Cover Detected by NDUI+ Dataset in Global Cities

NDUI+ demonstrates its utility for global urban analysis by capturing distinct urbanization patterns across diverse cities, including Delhi (India), Melbourne (Australia), and Dubai (UAE). Figure 5 illustrates these patterns, showcasing the dataset's ability to provide detailed, 30-meter resolution insights into the built environment. Higher NDUI values (red/blue) correspond to densely populated areas, while vegetated or non-urbanized zones exhibit lower values (white/yellow).

For example, the dataset reveals urban growth near Palm Jumeirah in Dubai between 1999 and 2010. Similarly, NDUI+ captures the dense urban sprawl of Delhi alongside its surrounding agricultural and natural areas. In Melbourne, the dataset delineates the compact urban core and the suburban expansion. These observations demonstrate NDUI+'s potential for global urban monitoring, enabling comparisons across cities with varying geographies, histories, and urbanization trajectories.

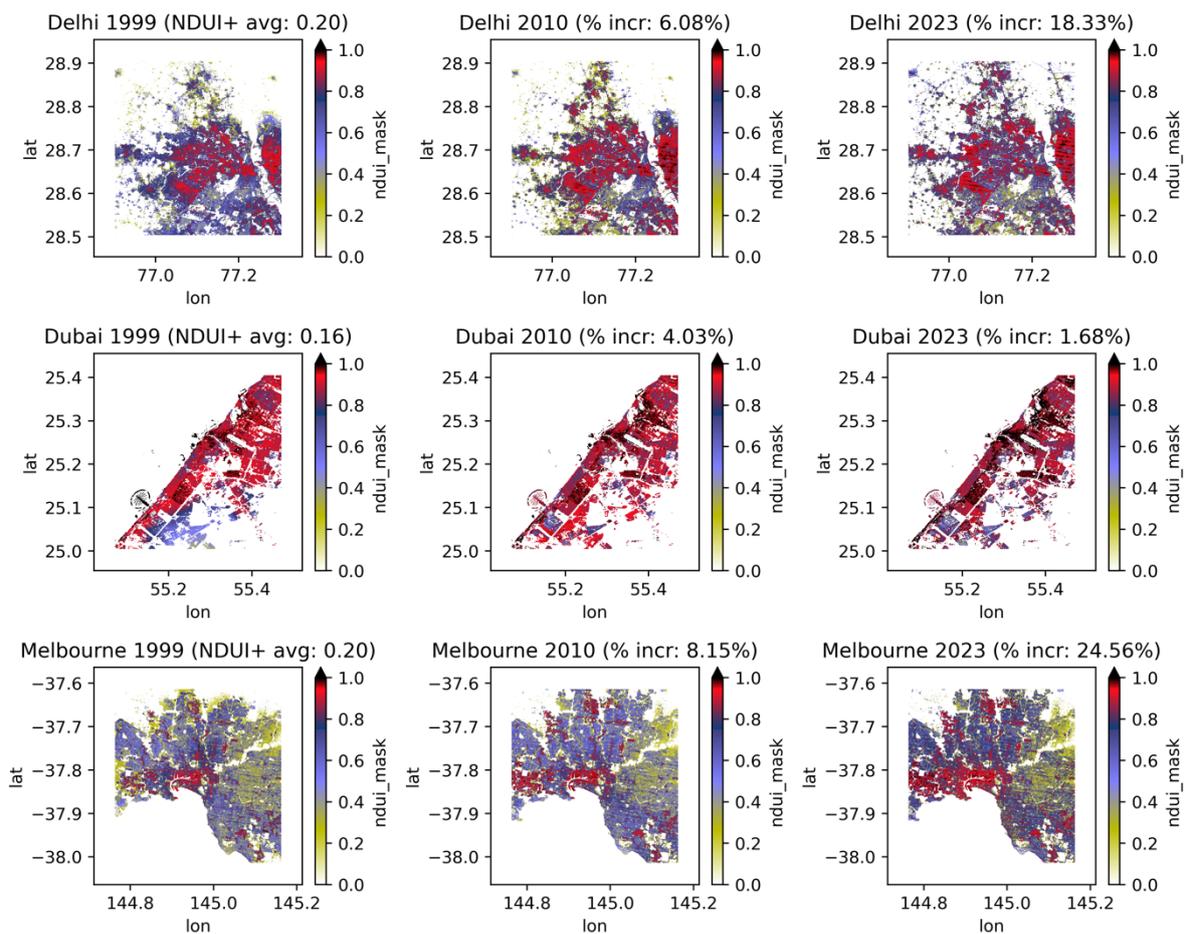

*Figure 5: Urban Expansion Detected by NDUI+ in global cities for the years 1999, 2010 and 2023. The figure subtitles show the absolute area average NDUI+ values in the first column and the percentage increase for the year 2010 relative to the year 1999 in the second column, and the percentage increase for the year 2023 relative to the year 2010 in the third column. Spatial maps represent absolute NDUI+ values.*

## 3. Discussion

This study successfully developed and validated the use of the AI-based Swin Transformer model to cross-calibrate inter-generational remote-sensing satellite data, resulting in the creation of the global, long-term, high-resolution NDUI+ dataset. The dataset spans 1999 to the present, providing annual temporal resolution and 30-meter spatial granularity. This approach addresses a critical gap in urban monitoring by overcoming the data discontinuity caused by the decommissioning of DMSP-OLS and ensuring compatibility with next-generation VIIRS data. The methodology offers a robust framework for tracking and evaluating urbanization trends over time, making it particularly suited for interdisciplinary research that bridges urban studies, climate science, and socio-environmental assessments.

The NDUI+ dataset's capability to capture urban boundaries while providing detailed insights into intra-urban structures enhances its utility for studies across local, regional, and global scales. By incorporating region-specific weight matrices, the methodology ensures adaptability to diverse urban geographies, from rapidly urbanizing cities in the Global South to mature urban systems in developed regions. This adaptability transcends geographical and disciplinary divides in urban scholarship. The simplicity and robustness of the proposed methodology also make it accessible for implementation in diverse research contexts, enabling policymakers, urban planners, and scientists to track urban growth, assess environmental impacts, and model socio-economic changes. For instance, the dataset can inform infrastructure planning, land use policy, and climate adaptation strategies by identifying patterns of urban expansion, densification, and environmental degradation.

The NDUI+ dataset uniquely provides continuity in urban information, overcoming challenges that have persisted in remote-sensing applications. It enables a detailed understanding of urbanization processes, including their spatial and temporal variability, and supports nuanced analyses of urban morphology, infrastructure, and socio-environmental interactions. The dataset's ability to integrate and harmonize data from multiple generations of satellite technology is a significant advancement, ensuring the sustainability of urban monitoring efforts over time. Moreover, NDUI+ demonstrates high fidelity in detecting small-scale urban features, such as roads and green roofs, making it valuable for urban climate zone studies and street-level urban heat island modeling. The dataset's granularity supports actionable urban planning, enabling fine-scale interventions in areas such as housing, transportation, and resource management. These applications provide scientifically rigorous insights with practical and policy relevance.

Despite its strengths, the study has certain limitations that point to avenues for future research. The dataset currently does not extend to periods before 1999, leaving gaps in historical urban analyses. Integrating older datasets such as Landsat could address this limitation, creating a longer time series for analyzing urban transitions over the past century. Additionally, NDUI+ does not yet classify urban areas into specific categories like roads, parks, or industrial zones, which could enhance its applicability for urban policy and socio-economic studies. Future work could focus on developing classification schemes that distinguish between different land-use types within urban areas. The methodology's reliance on VIIRS composites introduces potential challenges if VIIRS data production ceases. However, the framework is adaptable and could be applied to future generations of nighttime light data, although this remains to be empirically tested. Expanding the dataset to include multispectral or hyperspectral data could also provide richer insights into urbanization dynamics and their environmental impacts.

The NDUI+ dataset contributes to a deeper understanding of urbanization processes, addressing critical challenges in sustainability, resilience, and equity. By enabling detailed analyses of urban growth patterns and their environmental impacts, the dataset informs strategies for managing rapid urbanization, mitigating climate risks, and promoting equitable development. The ability to monitor urban dynamics at a fine scale supports the development of evidence-based policies that align with global sustainability goals, such as the UN's Sustainable Development Goals (SDGs). In **conclusion**, the NDUI+ dataset represents a significant step forward in urban science, providing a globally consistent, high-resolution tool for understanding and managing the complexities of urbanization. By addressing longstanding challenges in urban monitoring and offering practical applications for policymakers and researchers, this work fosters interdisciplinary collaboration and advancing knowledge at the intersection of urbanization, sustainability, and global change.

## 4. Methods

*4.1 Data Description*

We use three main satellite data products, the DMSP-OLS, VIIRS Nighttime Light images, and Landsat NDVI to calculate the NDUI+. For this, Google Earth Engine (GEE) was used to access the required satellite datasets. Additional details about these data and the process are discussed next.

*4.1.1 DMSP-OLS (Defense Meteorological Satellite Program - Operational Linescan System)*

The Operational Linescan System (OLS) of the Defense Meteorological Satellite Program (DMSP-OLS) has been compiling satellite data for a wide range of applications. The DMSP-OLS is a constellation of satellites that take images of Earth in both the visible and infrared spectrums at a spatial resolution of 30 arc seconds. The OLS captures worldwide nighttime low-light imagery (Elvidge et al., 1997). It has the capability to detect visible and near-infrared (VNIR) emission sources at night. It is especially helpful for viewing anthropogenic light sources like city lights and gas flares. Since the early 1970s, when the DMSP-OLS began providing continuous data, it has been possible to conduct long-term studies of numerous characteristics of Earth's surface. Examples include human settlement patterns, urban expansion and energy use. (Elvidge et al., 2009). Despite its popularity, DMSP-OLS sensors can get saturated in densely populated locations, making it difficult to track shifts (Elvidge et al., 2013). However, despite this caveat, due to its long-term, constant data collection and worldwide coverage, the DMSP-OLS remained an important resource for both scientific and policy-oriented applications (Huang et al., 2014). The GEE data repository contains the DMSP-OLS data for 1992-2014. This provides cloud-free composites made using all DMSP-OLS sensor data collected by US Air Force Weather Agency (AFWA) and processed by NOAA's National Geophysical Data Center (NGDC) for calendar years.

*4.1.2 VIIRS (Visible Infrared Imaging Radiometer Suite)*

The VIIRS sensor on the Suomi National Polar-orbiting Partnership (NPP) and the NOAA-20 satellites collects the Visible Infrared Imaging Radiometer Suite (VIIRS) Nighttime Lights Data. This information allows observation and measurement of nocturnal light emissions, as well as images and measurements of Earth's atmosphere, seas, and land surfaces (Elvidge et al., 2013). The VIIRS sensor has a resolution of 375-750 m and captures data in 22 spectral bands from the visible to the long-wave infrared. The Day/Night Band (DNB) is a standout feature since it can pick up faint signals during the night. In order to give "visible" pictures when sunlight is limiting, the DNB employs a mix of moonlight, airglow, zodiacal light, stars, and anthropogenic light sources (Miller et al., 2013). Urbanization, population change, economic activity, and even power outages and natural hazards have been tracked with VIIRS (Levin et al., 2020). For instance, researchers have utilized VIIRS data to observe shifts in nighttime light patterns to monitor the rehabilitation of communities following natural catastrophes (Roman et al., 2019). Additionally, urban ecosystems research has also made use of VIIRS's Nighttime Lights Data. Many animals are impacted by artificial lights. VIIRS NTL data has been studying their behavior and migration patterns induced by light pollution for years. (Gaston et al., 2012).

*4.1.3 Normalized Difference Vegetation Index (NDVI) derived from Landsat*

The Normalized Difference Vegetation Index (NDVI) has been widely used metric in remote sensing studies. It is calculated using different sensors such as MODIS and Landsat. NDVI provides critical insights into terrestrial vegetation health and productivity, and is calculated as:

$$NDVI = \frac{\text{NIR} - \text{RED}}{\text{NIR} + \text{RED}}$$

where, NIR represents near-infrared reflectance and RED represents red light reflectance. By measuring the difference between these spectral bands, NDVI indicates the chlorophyll content (greenness) of vegetation cover (Rouse et al., 1974, Ghosh et al., 2020). This index ranges from -1 to 1, with higher values indicating more vigorous plant metabolism.

*4.1.4 Normalized Difference Urban Index (NDUI)*

The Normalized Difference Urban Index (NDUI) as discussed is an analog to NDVI but for urban landscape (Zhang et al. 2015). The NDUI matrix utilizes data from the Defense Meteorological Satellite Program's Operational Line-Scan System (DMSP-OLS) and Landsat NDVI. The equation is as follows:

$$NDUI = \frac{NTL - NDVI}{NTL + NDVI}$$

In the above, NTL represents the rectified DMSP-OLS nighttime image. The core concept of NDUI is that the vegetation profiles and luminosity intensity passing through urban cores present an inverse relationship i.e. vegetation density decreases while luminosity increases from rural areas to the urban core (Zhang et al., 2015).

NDUI has been a useful dataset especially for land models in weather and climate studies (Demuzere et. al., 2019, Huang et. al., 2019). The transition and interpretation of the data from vegetated to urban grids through NDVI and NDUI distinction added to its functional popularity (Peng et. al., 2018). Furthermore, the NDUI data matrix was also found to be robust in its ability to capture urban features. Figure S3 compares the NDUI matrix with a Google Earth image for the area around The University of Texas at Austin. Most of the region is urbanized with limited greenspaces, with the exception of the football stadium, and dedicated park. Also notable are the urban developments in the areas. It can be seen that NDUI can capture urban and non-urban features with high fidelity. In the figure, the Google Earth image shows clear urban features like roads, buildings, a stadium, and park areas. The NDUI data, with a 30-meter resolution, effectively distinguishes built areas from vegetated ones.

*4.2 Methodological Framework*

This section outlines the methodology adopted for developing the VIIRS, DMSP-OLS fused NDUI dataset, referred henceforth as NDUI+ to distinguish from classical NDUI. There are four related elements: (i) Cross-calibration between VIIRS and DMSP-OLS, (ii) Deep learning model selection for the cross-calibration, (iii) Generation of global DMSP-OLS-Like dataset from VIIRS and (iv) Calculation of NDUI+ through data fusion. Each of these aspects are discussed next.

*4.2.1 Cross-Calibration between VIIRS and DMSP-OLS*

One of the challenges of developing and maintaining long-term satellite datasets is the issue of discontinuity of satellite missions and technological upgrades of sensors. Remote sensing satellites have a limited operational lifespan, and as newer missions are launched, older satellites are decommissioned or become non-operational. This leads to a disruption in the continuity of data collection, making it difficult to establish consistent time-series datasets for long-term analysis. Additionally, advancements in sensor technology often result in the deployment of satellites with improved capabilities and higher-resolution sensors. While this is beneficial for obtaining more detailed information, it also introduces a change in the sensor characteristics, making it challenging to compare data collected by different generations of satellites. Therefore, to maintain data continuity, a normalization process is typically followed.

There have been prior attempts to create DMSP-OLS like dataset to extend the time series of the original DMSP Nighttime Lights data through various cross-sensor calibration approaches (e.g. Ghosh et al., 2021; X. Li et al., 2013; Nechaev et al., 2021; Tu et al., 2020; Zheng et al., 2019). They used different approaches such as pseudo-invariant functions (PIF) paradigm or regression models. However, the problem with most methods is that they assume a linear or polynomial relationship between DMSP and VIIRS radiances, which is not always the case due to the different resolution, dynamic range, and saturation effects of the sensors (X. Li et al., 2017).

Studies such as Ghosh et al. (2021) extended the series of DMSP from 2013 to 2019 based on a Convolutional Neural Network (CNN) based model (Residual U-Net) and is discussed in Nechaev et al., (2021). Since the primary (F18) satellite of DMSP stopped collecting usable nighttime data from the beginning of 2014, their work took F15 and F16 satellites of DMSP that have been collecting pre-dawn data, and F18 and F15 satellite images for early-evening period. This was aided by the satellite sliding from day/night orbit to dawn/dusk orbit from 2012 onwards and calibrate it with VIIRS mid-night data to study the diurnal pattern of nighttime lights. Following this method, the authors were able to extend the data timeframe from 2013 to 2019 (Nechaev et al., 2021). This method relied on data augmentation and used the data for the year DMSP-OLS and VIIRS NTL maps were available for the network training. This has two limitations. First, it is not possible anymore since the orbits of the DMSP-OLS satellites have further degraded and not enough nighttime data is collected to make annual nighttime lights. Second, there is only limited year of overlap data.

*4.2.2 Deep learning Model Selection for the Cross-Calibration*

In this study, we use the UT-ClimateDownscaleSuite (UT-CDS) developed at University of Texas at Austin to test different deep-learning models (Singh et al 2024). The UT-CDS provided a convenient platform to choose the optimal model for the cross-calibration task. A total of 17 potential deep learning models were tested. These include. Super-Resolution Convolutional Neural Network (SRCNN), Fast SRCNN (FSRCNN), Cascading Residual Network for Super-Resolution (Carn-M), Laplacian Pyramid Super-Resolution Network (LapSR), Balanced Two-Stage Residual Networks (BTSRN), Fast, Accurate and Lightweight Super-Resolution models (FalSR-a, FalSR-b), Optical Image Stabilization Based Super Resolution (OISSR), Multi-Scale Deep Super-Resolution (MDSR), Second-order Attention Network (SAN), very deep Residual Channel Attention Networks (RCAN), Dynamic Local and Global Self-Attention Network (DLGSANet), Dual Prior Modulation Network (DPMN), Spatially-Adaptive Feature Modulation Network (SAFMN), Dense Prediction Transformers (DPT), Attention in Attention Network (A2N) and Swin Transformer. The Swin Transformer model (Liang et al., 2021) was chosen to fuse the DMSP-VIIRS as it yielded the highest peak signal to noise ratio in the test data during training.

The Swin Transformer builds upon the Vision Transformer architecture (ViT) (Dosovitskiy et al., 2021), which applies the transformer architecture initially developed for natural language processing tasks to computer vision. The ViT model segments images into fixed-size data patches, linearly embeds these patches, and then applies a sequence of transformer layers. The Swin Transformer enhances this approach by using a hierarchical structure that applies self-attention across local windows of image patches and then shifts these windows across layers. This strategy allows the model to capture both local and global image contexts effectively while maintaining computational efficiency. In other words, ViT model processes images by segmenting them into smaller patches and analyzing each patch individually. This approach allows the model to capture fine-grained details within each segment while maintaining an understanding of the overall image structure through advanced computational techniques. The Swin Transformer builds upon this concept by hierarchically organizing these patches into non-overlapping local windows and subsequently integrating information across different windows. This hierarchical strategy enables the model to efficiently capture both local and global contextual information within an image, thus providing a more comprehensive understanding of the visual content. This approach has been tested and yielded good success in various studies in computer vision and geoscience fields (e.g. Dai et. al., 2024, He et al., 2022). By simulating interactions at multiple scales, the Swin Transformer achieves notable performance improvements across various applications, and is highlighted for its versatility and computational efficiency (Y. Li et al., 2022; Liu et al., 2022).

*4.2.3 Generation of Global DMSP-OLS-Like Dataset from VIIRS*

Since the data availability of DMSP-OLS and VIIRS overlapped in 2012, the data from this year is selected as reference and calibration to harmonize the datasets. The global data was subdivided into 5x5 degree grid boxes for training local Swin Transformer models focused on the respective regions. Globally cities information with respective latitude, longitude information were used to guide the Swin Transformer model to generate a weight matrix for the cities all over the world from the input historical DMSP-OLS image collections. The model was then programmed to use those weight matrices to calibrate the input VIIRS image collections over the same geographical areas and harmonize VIIRS data for subsequent years. This harmonized VIIRS data acts as DMSL-OLS-like data.
Figure S4 below shows comparative scenes from DMSP, VIIRS, and the produced "calibrated" DMSP-OLS-like data developed from VIIRS using Swin Transformer. The result was validated against a similar DMSP-OLS-like data (available for a very limited time period) that was available from Ghosh et al (2021).

*4.2.4 Calculation of NDUI+ through Data Fusion*

The calibrated VIIRS (calibVIIRS) dataset was used as input nighttime light data and combined with Landsat NDVI to calculate annual NDUI for the year 2012 and subsequent years NDUI+. The equation used for the calculation is as follows.

$$NDUI+ = \frac{CalibVIIRS - NDVI}{CalibVIIRS + NDVI}$$

where, NDUI+ is the long-term NDUI from this study, CalibVIIRS is the calibrated VIIRS data similar to DMSP-OLS, and NDVI is Normalized Difference Vegetation Index. The output was then combined with the DMSP-OLS-derived NDUI matrix, creating the new NDUI+ (30 m) dataset. Moreover, water and other non-urban pixels can introduce noise and false high NDUI. We enhance the NDUI+ data by mitigating the noise and non-urban data. For this, raw NDUI+ was refined by filtering out non-urban areas for each year based on the Dynamic World built up data available since 2015 (Brown et al., 2022). For this, pre-2015 built up used to filter raw NDUI series is taken as that for the year 2015. Figure S5 illustrates the overall methodological framework.

**Acknowledgment:** The authors would like to thank the anonymous reviewers for their valuable comments and feedback on the quality enrichment of the manuscript.

**Funding:** This study has been supported through the NASA Interdisciplinary Research in Earth Science (IDS) program (Grant No. NNH19ZDA001N-IDS). The content of this report is solely the responsibility of the author and does not necessarily represent the official views of the funding agency.

**Conflict of Interest:** The Authors declare no conflict of interest.

**Data Availability:** The DMSP-OLS data can be accessed here: https://developers.google.com/earth-engine/datasets/catalog/NOAA_DMSP-OLS_NIGHTTIME_LIGHTS, the VIIRS NTL data can be accessed here: https://developers.google.com/earth-engine/datasets/catalog/NOAA_VIIRS_DNB_MONTHLY_V1_VCMCFG, The generated annual NDUI+ data for the year 1999-2023 can be accessed here: https://zenodo.org/records/10799652.

**Code Availability:** The codes used for this study are available here: https://github.com/texuslab/NDUI+.

Elvidge, C. D., Baugh, K., Zhizhin, M., & Hsu, F. C. (2013). Why VIIRS data are superior to DMSP for mapping nighttime lights. Proceedings of the Asia-Pacific Advanced Network, 35, 62-69.

C. D. Elvidge, K. Baugh, M. Zhizhin, F. C. Hsu, and T. Ghosh, "VIIRS night-time lights," International Journal of Remote Sensing, vol. 38, pp. 5860–5879, 2017.

Gaston, K.J., Davies, T.W., Bennie, J., & Hopkins, J. (2012). REVIEW: Reducing the ecological consequences of night-time light pollution: options and developments. Journal of Applied Ecology, 49(6), 1256–1266.

Ghosh, S., Bandopadhyay, S., & Sánchez, D. A. C. (2020). Long-Term Sensitivity Analysis of Palmer Drought Severity Index (PDSI) through Uncertainty and Error Estimation from Plant Productivity and Biophysical Parameters. The 1st International Electronic Conference on Forests—Forests for a Better Future: Sustainability, Innovation, Interdisciplinarity, 57. https://doi.org/10.3390/IECF2020-07956

Ghosh, T., Baugh, K. E., Elvidge, C. D., Zhizhin, M., Poyda, A., & Hsu, F.-C. (2021). Extending the DMSP Nighttime Lights Time Series beyond 2013. Remote Sensing, 13(24), 5004. https://doi.org/10.3390/rs13245004

Guo, W., Lu, D., & Kuang, W. (2017). Improving Fractional Impervious Surface Mapping Performance through Combination of DMSP-OLS and MODIS NDVI Data. Remote Sensing, 9(4), Article 4. https://doi.org/10.3390/rs9040375

Global Economic Outlook. (n.d.). The Conference Board. Retrieved July 12, 2024, from https://www.conference-board.org/topics/global-economic-outlook

Guzder-Williams, Brookie, Eric Mackres, Shlomo Angel, Alejandro M. Blei, and Patrick Lamson-Hall. "Intra-urban land use maps for a global sample of cities from Sentinel-2 satellite imagery and computer vision." Computers, Environment and Urban Systems 100 (2023): 101917.

He, X., Zhou, Y., Zhao, J., Zhang, D., Yao, R., & Xue, Y. (2022). Swin transformer embedding UNet for remote sensing image semantic segmentation. IEEE Transactions on Geoscience and Remote Sensing, 60, 1-15.

Huang, X., Wang, C., & Lu, J. (2019). Understanding the spatiotemporal development of human settlement in hurricane-prone areas on the US Atlantic and Gulf coasts using nighttime remote sensing. Natural Hazards and Earth System Sciences, 19(10), 2141-2155.

Huang, X., Li, J., Yang, J. et al. 30 m global impervious surface area dynamics and urban expansion pattern observed by Landsat satellites: From 1972 to 2019. Sci. China Earth Sci. (2021). https://doi.org/10.1007/s11430-020-9797-9

**Supplementary**

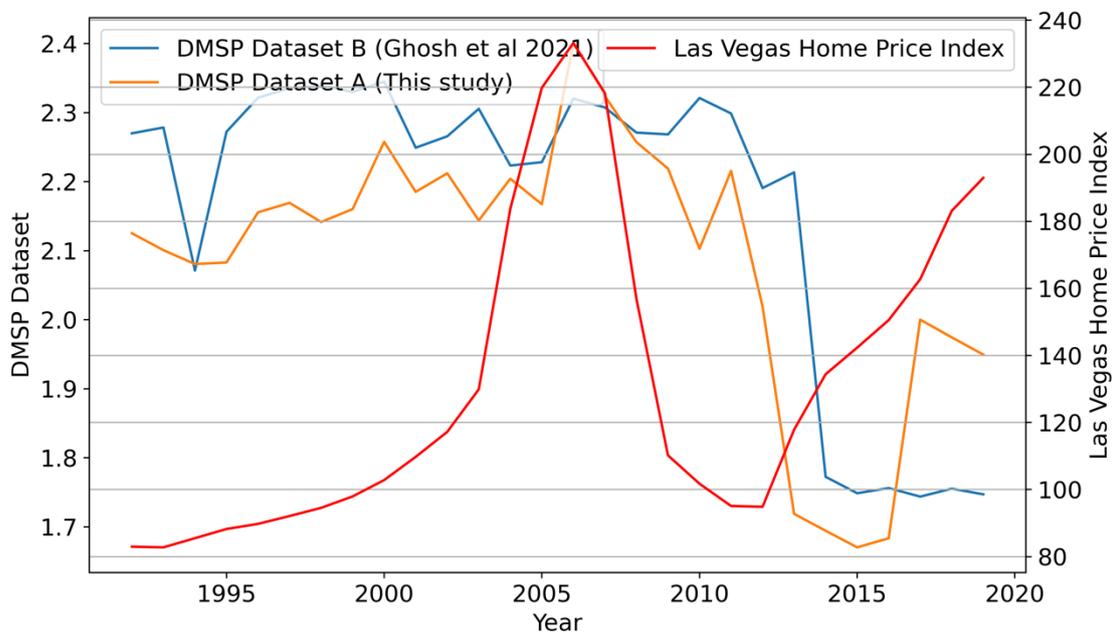

*Figure S1: Comparison between DMSP-OLS-like nighttime light data generated Dataset A over Las Vegas, USA along with Las Vegas Home Price Index. (Home price index data Source: https://fred.stlouisfed.org/series/LVXRNSA). Night light is indicative of urbanization and human activity.*

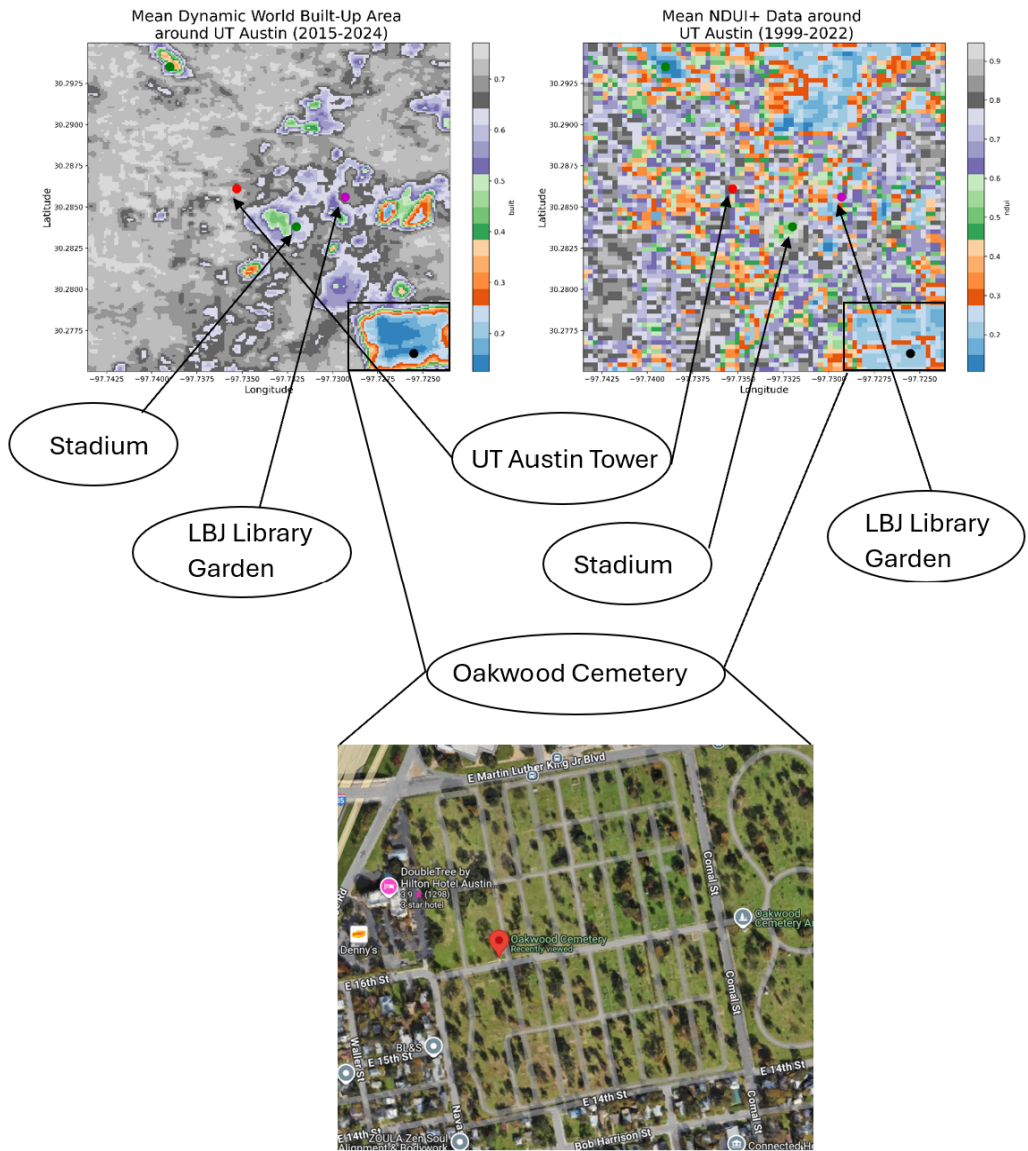

*Figure S2: Comparative scenes from Dynamic World and NDUI+ over UT Austin, USA. NDUI+ being able to detect roads within the Oakwood Cemetery when the Dynamic World failed.*

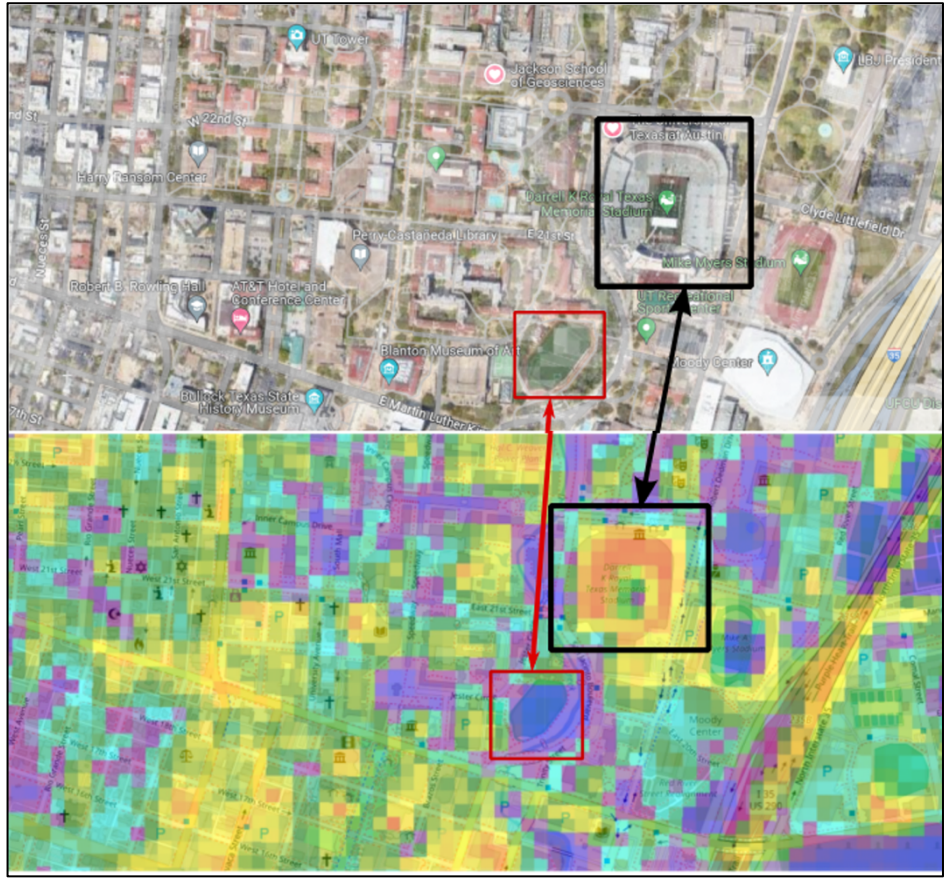

*Figure S3: Comparison between Google Earth image and the Normalized Difference Urban Index (NDUI) for the year 2012. Yellow and dark red colors indicate high NDUI values, while purple and blue represent low NDUI values. The marked A box around the stadium highlights how high urban concentration areas (stadium, sitting areas etc.) have higher NDUI values (yellow/red), while the green field inside shows lower values (blue/violet), indicating vegetation. Similarly, the box B around the park shows it mostly in blue/violet, highlighting its vegetative nature. The NDUI also clearly outlines urban roads, demonstrating its usefulness in analyzing urban areas. This NDUI data has been reproduced using the original DMSP-OLS product following Zhang et al., (2015)*

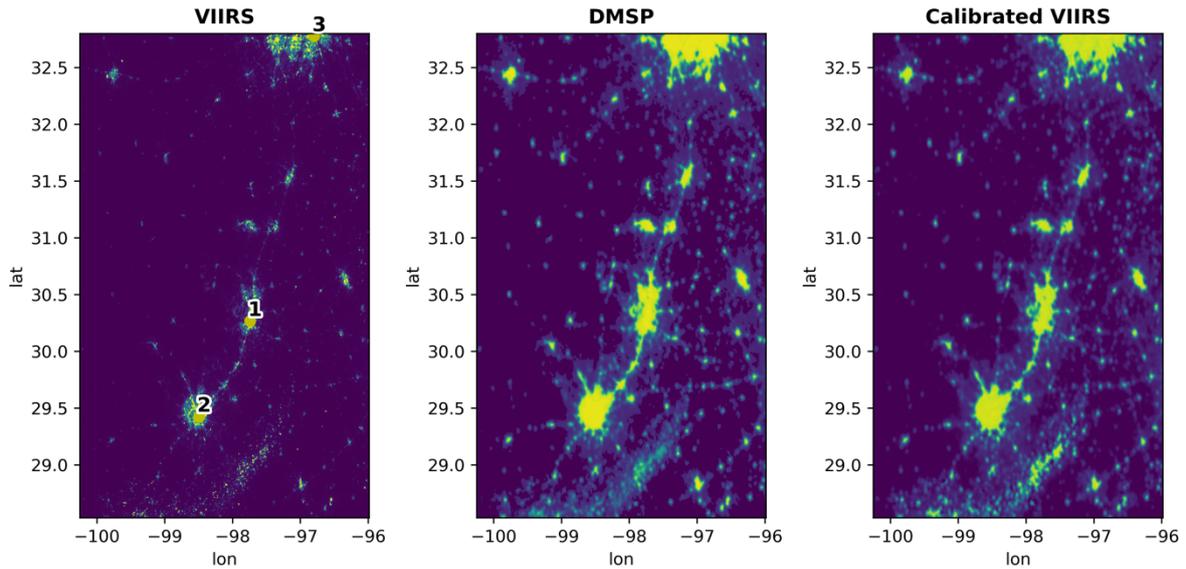

*Figure S4: Comparative scenes from VIIRS, DMSP, and the calibrated VIIRS (Simulated DMSP-OLS) over San Antonio-Austin area in Texas, USA. The cities are marked as 1- Austin, 2- San Antonio. Part of city of Dallas footprint is also captured in this image at the top edge marked as 3 in the figure.*

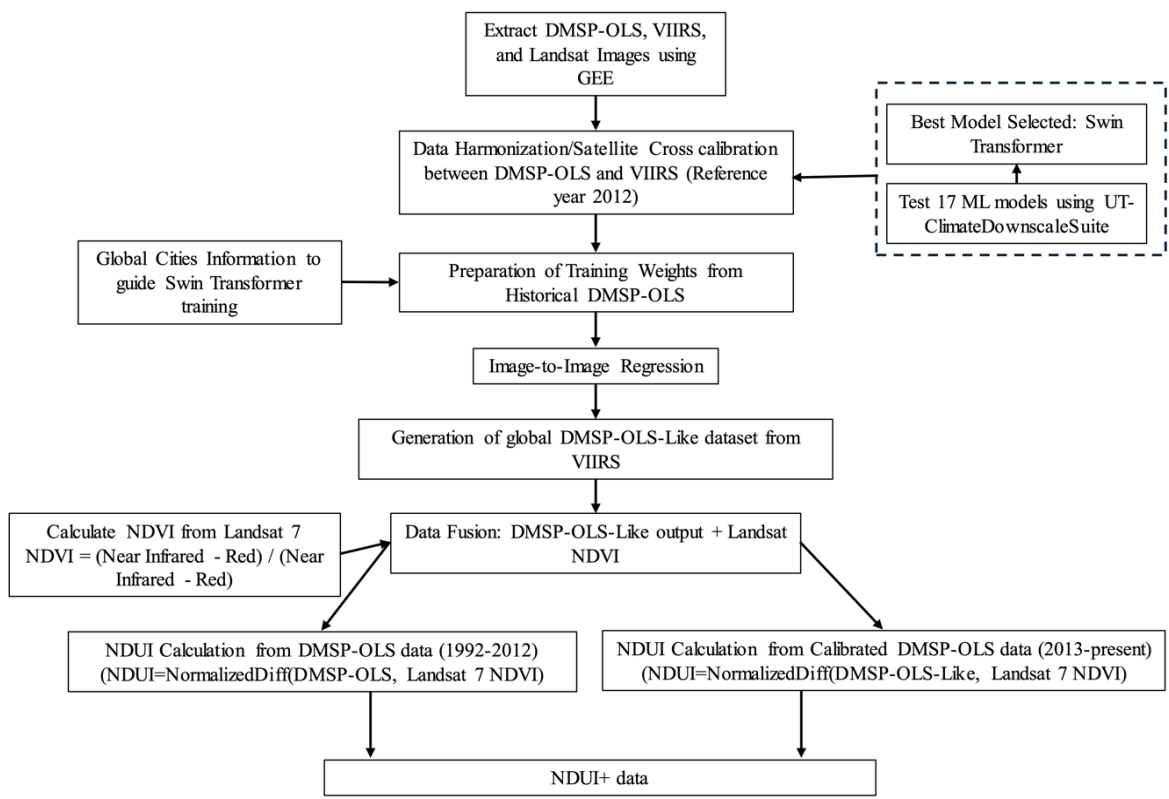

*Figure S5: Overall methodological framework describing the algorithm used to develop the NDUI+ dataset*

25